\documentclass[conference]{IEEEtran}

\usepackage{amsmath,amsfonts,bm}

\def\eqref#1{equation~\ref{#1}}

\def\1{\bm{1}}

\DeclareMathAlphabet{\mathsfit}{\encodingdefault}{\sfdefault}{m}{sl}
\SetMathAlphabet{\mathsfit}{bold}{\encodingdefault}{\sfdefault}{bx}{n}

\usepackage{hyperref}
\usepackage{url}
\usepackage{amsmath, bm}

\usepackage{multirow}
\usepackage{arydshln}
\usepackage{booktabs}
\usepackage{pgfplots} 
\pgfplotsset{compat=1.17}

\begin{document}

\title{Variational Diffusion Channel Decoder}

\author{
    \IEEEauthorblockN{Chengwei Zhang, Yifan Du, Siyu Liao}
    \IEEEauthorblockA{\textit{The School of Integrated Circuits} \\
    \textit{Sun Yat-sen University} \\
    Shenzhen, China \\
    \{17858869215@163.com, duyf23@mail2.sysu.edu.cn, liaosy36@mail.sysu.edu.cn\}}
}

\maketitle

\begin{abstract}
Neural channel decoder, as a data-driven channel decoding strategy, has shown very promising improvement on error-correcting capability over the classical methods. However, the success of those deep learning-based decoder comes at the cost of drastically increased model storage and computational complexity, hindering their practical adoptions in real-world time-sensitive resource-sensitive communication and storage systems. To address this challenge, we propose an efficient variational diffusion model-based channel decoder, which effectively integrates the domain-specific belief propagation process to the modern diffusion model. By reaping the low-cost benefits of belief propagation and strong learning capability of diffusion model, our proposed neural decoder simultaneously achieves very low cost and high error-correcting performance. Experimental results show that, compared with the state-of-the-art neural channel decoders, our model provides a feasible solution for practical deployment via achieving the best decoding performance with significantly reduced computational cost and model size. 
\end{abstract}


\section{Introduction}
Channel coding has served as the fundamental and critical mechanism in numerous modern communication and storage systems and applications, such as 5G, Wi-Fi, Starlink, optical networking, solid-state drive and hard disk drive. By providing error correction functionality, channel coding aims at protecting information from various corruptions (e.g., noise) incurred by data transmission. To that end, most channel codes are designed by adding extra redundant bits to help detect and recover the original information after noisy transmission.

To date, most of the commercially adopted channel codes are linear block codes, which can be optimally decoded using maximum likelihood (ML) decoding process. 
However, ML decoding, which can be mathematically modeled as searching for the closest lattice point in high dimensional space \cite{gowaikar2007statistical}, is very expensive and computationally prohibitive. 
In practice, a more feasible and practical channel decoding solution is to use belief propagation (BP) algorithm \cite{su202258}, which can achieve exact optimum results in the tree-structured factor graph. However, in the context of channel coding, factor graphs constructed from the parity check matrix of modern channel codes are often cyclic, making the BP decoding results suboptimal \cite{yedidia2005constructing}.   

\textbf{Existing Neural Channel Decoders.} Recent works leverage deep learning to improve decoding performance. \cite{nachmani2016learning,nachmani2019hyper} add learnable weights to BP messages, yet remain constrained to the BP graph structure. \cite{bennatan2018deep} bypass this by reformulating decoding as noise prediction with syndrome as auxiliary input, enabling flexible model design. Building on this, \cite{choukroun2022error} introduce a transformer-based decoder \cite{vaswani2017attention}, later extended to a diffusion-based model \cite{DBLP:conf/iclr/ChoukrounW23}.

\textbf{High-Complexity Challenges of SOTA Neural Decoders.} While today's deep learning-based decoding algorithms achieve outstanding error-correcting performance, they suffer from high computation and storage costs that severely hinder practical deployment. Channel coding applications demand real-time, low-power processing—e.g., decoding latency in 5G is limited to millisecond level \cite{parvez2018survey, rico2020survey}—while most deployments occur on resource-constrained mobile devices. Consequently, unless model cost is properly addressed, widespread adoption of neural channel decoders remains infeasible.

\textbf{Technical Contributions.} We propose VCDC, built on the variational diffusion model (VDM) \cite{kingma2021variational}. Unlike \cite{DBLP:conf/iclr/ChoukrounW23}, we integrate BP into the diffusion process: since BP operates on LLR inputs, VDM's unconstrained $\alpha_t$/$\sigma_t$ parameterization naturally accommodates the AWGN channel as the forward process. Our contributions are: (1) the first integration of BP into a diffusion model framework for channel decoding; (2) a neural decoder achieving ultra-low complexity while matching or surpassing state-of-the-art BER performance.

\section{Related Works}
Neural channel decoders broadly fall into two categories: BP-based and general neural network-based models. BP-based models augment the message-passing structure with learnable weights \cite{nachmani2016learning, lugosch2017neural, liang2018iterative, nachmani2019hyper, liao2021doubly}, improving performance while remaining constrained to the BP graph structure. General neural network-based models lift this constraint: \cite{gruber2017deep} propose a dense-layer decoder for short codes; \cite{bennatan2018deep} recast decoding as noise prediction using syndrome as auxiliary input and a binary-input symmetric-output channel framework, enabling flexible model design. Building on this, \cite{choukroun2022error} apply the transformer architecture \cite{vaswani2017attention}, later extended to a DDPM-based model \cite{DBLP:conf/iclr/ChoukrounW23} where the AWGN channel serves as the unscaled forward diffusion and decoding steps are bounded by parity-check count.

High complexity remains a central challenge for diffusion-based decoders. Several orthogonal methods address this—reduced sampling steps \cite{DBLP:conf/iclr/SongME21}, faster ODE solvers \cite{lu2022dpm}, and knowledge distillation \cite{DBLP:conf/iclr/SalimansH22}—all of which are potentially applicable to our approach for further speedup.

\section{Background}
\subsection{Channel Coding}
An $(N,K)$ channel code is defined by generator matrix $\mathbf{G}\in\{0,1\}^{K\times N}$ and parity check matrix $\mathbf{H}\in\{0,1\}^{(N-K)\times N}$. A $K$-bit message $\mathbf{m}^b$ is encoded as $\mathbf{x}^b=\mathbf{m}^b\mathbf{G}\in\{0,1\}^N$ and transmitted over a noisy channel; the receiver recovers $\mathbf{m}^b$ from the corrupted signal via systematic decoding \cite{lin2004error}.

In the memoryless AWGN channel, the transmitted output $\mathbf{y}_s\in\mathbb{R}^N$ is simulated by $\mathbf{y}_s=\mathbf{x}+w_s\mathbf{\xi}$ with $\mathbf{\xi}\sim \mathcal{N}(0, \mathbf{I})$.
Here, $\mathbf{x}\in\{-1,1\}^N$ is the bipolar representation computed from $\mathbf{x}=1-2\mathbf{x}^b$. 
The $w_s$ is determined by code rate $r=K/N$ and the channel signal-to-noise ratio (CSNR) in $s$-dB, i.e., $w_s=1/\sqrt{2\frac{K}{N}10^{s/10}}$.
While there are many different decoding algorithms, belief propagation (BP) has achieved tremendous success in channel codes. 
It treats the parity check matrix as a factor graph, and belief messages are iteratively propagated and updated over the graph. 

More specifically, for a given parity check matrix, there are $N$ variable nodes and $N-K$ check nodes in the associated factor graph. 
First, define a compact representation of transmitted messages using log-likelihood ratio (LLR):
\begin{equation}
\begin{aligned}
l_v &= \text{LLR}(\mathbf{y}_s^v) \\
    &= \log\frac{p(\mathbf{y}_s^v|\mathbf{x}_v=1)}{p(\mathbf{y}_s^v|\mathbf{x}_v=-1)} \\
    &= \frac{2}{w_s^2}\mathbf{x}_v+\frac{2}{w_s}\mathbf{\xi}_v,
\end{aligned}
\end{equation}
where $\mathbf{y}_s^v$ is the $v$-th value of $\mathbf{y}_s$.
LLR-based belief propagation estimates the $v$-th bit of $\mathbf{x}$ by computing:
\begin{equation}
\label{eq:bp}
\begin{aligned}
u_{c\rightarrow v} &= 2\operatorname{arctanh}\!\bigg[\prod_{v'\in M(c)\backslash v}
    \tanh\!\left(\frac{u_{v'\rightarrow c}}{2}\right)\bigg], \\
u_{v\rightarrow c} &= l_v + \sum_{c'\in N(v)\backslash c}u_{c'\rightarrow v}, \\
s_v &= l_v + \sum_{c'\in N(v)}u_{c'\rightarrow v},
\end{aligned}
\end{equation}
where $M(\cdot)$ and $N(\cdot)$ denote the neighboring variables and check nodes, respectively.
$\mathbf{x}_v$ can be determined by the sign of $s_v$. 
The expensive computation cost of hyperbolic tangent function in $u_{c\rightarrow v}$ can be simplified into more implementation-friendly operations \cite{hu2001efficient}:
\begin{equation}
\label{eq:min_sum}
\begin{aligned}
u_{c\rightarrow v} &= \min_{v'\in M(c)\backslash v}|u_{v'\rightarrow c}| \\
&\quad \times \prod_{v'\in M(c)\backslash v}\operatorname{sign}{(u_{v'\rightarrow c})},
\end{aligned}
\end{equation}
where the $\operatorname{sign}(\cdot)$ function returns the sign of input. 

\subsection{Diffusion Models}
DDPMs \cite{ho2020denoising,sohl2015deep} model complex distributions by gradually corrupting data in a forward Markov chain and learning a reverse denoising process. Let $\mathbf{x}_0\sim q(\mathbf{x}_0)$; the forward process adds Gaussian noise at each timestep:
\begin{equation}
\begin{aligned}
q(\mathbf{x}_T, \dots, \mathbf{x}_1) &= \prod_{t=1}^T q(\mathbf{x}_t|\mathbf{x}_{t-1}), \\
\mathbf{x}_t &= \sqrt{1-\beta_t}\mathbf{x}_{t-1} + \sqrt{\beta_t}\mathbf{\xi}_t, \\
&\quad \mathbf{\xi}_t \sim \mathcal{N}(0, \mathbf{I}),
\end{aligned}
\end{equation}
where $\beta_t$ for all $t$ are predefined hyper-parameters.
With $\alpha_t=1-\beta_t$ and $\bar{\alpha}_t=\prod_{s=1}^t \alpha_t$, 
we can have $\mathbf{x}_t=\sqrt{\bar{\alpha}_t}\mathbf{x}_0+\sqrt{(1-\alpha_t)}\xi$.
The reverse process learns the Gaussian distribution $p_\theta(\mathbf{x}_{t-1}|\mathbf{x}_t)=\mathcal{N}(\mathbf{\mu}_\theta(\mathbf{x}_t, t), \sigma_t^2\mathbf{I})$ to form the Markov chain:
\begin{equation}
\begin{aligned}
p(\mathbf{x}_T, \dots, \mathbf{x}_0) &= p(\mathbf{x}_T)\prod_{t=1}^T p(\mathbf{x}_{t-1}|\mathbf{x}_{t}), \\
\mathbf{x}_{t-1} &= \mathbf{\mu}_\theta(\mathbf{x}_t, t) + \sigma_t\xi,
\end{aligned}
\end{equation}
where $\theta$ describe model parameters and $\sigma_t$ is a function of $\beta_t$ for all $t$. 
The learning objective is simplified from the evidence lower bound (ELBO) to the KL divergence between $q(\mathbf{x}_{t-1}|\mathbf{x}_0, \mathbf{x}_t)$ and $p_\theta(\mathbf{x}_{t-1}|\mathbf{x}_t)$, which has an analytical form due to their Gaussian nature:
\begin{equation}
\begin{aligned}
\theta^* &= \operatorname*{arg\,min}_\theta \sum_{t > 1} \mathbb{E} \Bigg[ \frac{1}{2\sigma_t^2} \bigg\| \\
&\quad \tfrac{\sqrt{\bar{\alpha}_{t-1}}\beta_t}{1-\bar{\alpha}_t}\mathbf{x}_0
    + \tfrac{\sqrt{\alpha_t}(1-\bar{\alpha}_{t-1})}{1-\bar{\alpha}_t}\mathbf{x}_t \\
&\quad - \mathbf{\mu}_\theta(\mathbf{x}_t, t) \bigg\|^2 \Bigg],
\end{aligned}
\end{equation}
where efficient training optimizes at a random timestep via stochastic gradient descent.
Since $\mathbf{x}_t$ can be expressed in terms of $\mathbf{x}_0$, the model $\mu_\theta(\mathbf{x}_t,t)$ can equivalently predict $\mathbf{x}_0$ or noise $\mathbf{\xi}_t$; DDPMs adopt the noise-prediction parameterization. 

\section{Variational Diffusion Channel Decoder}

It can be noticed that the parameterization in DDPMs is specially designed, such as the relationship between $\alpha$ and $\beta$ and  between $\mathbf{x_t}$ and $\mathbf{x}_0$. 
Although this specific design results in clear formulations for training and generation, such constraints are not realistic in channel coding, especially when describing the AWGN channel from the perspective of forward diffusion process.  
In this section, we propose an efficient decoding method using a flexible variational diffusion models (VDMs) framework \cite{kingma2021variational}, i.e., variational diffusion channel decoder (VCDC). 

\subsection{AWGN and Forward Process}
Different from DDPMs, VDMs generalize the mean and variance setting in the forward diffusion process. 
It enables the flexible Gaussian transition $q(\mathbf{x}_t|\mathbf{x}_0)=\mathcal{N}(\alpha_t\mathbf{x}_0, \sigma_t^2\mathbf{I})$, without constraints on the relation between  $\alpha_t$ and $\sigma_t$. 
We find such flexibility better help describe the AWGN channel as the forward diffusion process than DDPMs, especially with inputs in LLR format.
For different AWGN channels in $T$ different CSNRs, we define the forward diffusion process by transmitting bipolar codeword $\mathbf{x}$ across these  channels and the reverse process by denoising the transmitted messages at $t$-th CSNR to recover $\mathbf{x}$.

More specifically, let $\mathbf{z}_s=\text{LLR}(\mathbf{y}_s)$, and we have the distribution $q(\mathbf{z}_s|\mathbf{x})=\mathcal{N}(\frac{2}{w_s^2}\mathbf{x}, \frac{4}{w_s^2}\mathbf{I})$, where $\alpha_s=\frac{2}{w_s^2}$ and $\sigma_s=\frac{2}{w_s}$. 
Let $\mathbf{z}_t$ be another channel in $t$-dB CSNR.
According to VDM, the forward transition probability from $s$-dB channel to $t$-dB channel is:
\begin{equation}
\begin{aligned}
q(\mathbf{z}_t|\mathbf{z}_s) &= \mathcal{N}(\alpha_{t|s}\mathbf{z}_s, \sigma^2_{t|s}\mathbf{I}) \\
&= \mathcal{N}\left(\frac{\alpha_t}{\alpha_s}\mathbf{z}_s, (\sigma_t^2-\alpha^2_{t|s}\sigma_s^2)\mathbf{I}\right),
\end{aligned}
\end{equation}
where $\sigma^2_{t|s}$ should be positive since $s$ and $t$ are different:
\begin{equation}
\begin{aligned}
\sigma^2_{t|s} &= \sigma_t^2-\alpha^2_{t|s}\sigma_s^2 \\
&= \left(\frac{2}{w_t}\right)^2 - \left(\frac{w_s^2}{w_t^2}\right)^2\!\left(\frac{2}{w_s}\right)^2 \\
&\quad > 0 \implies w_t > w_s \implies t < s.
\end{aligned}
\end{equation}
It is worth noticing that $s$ describes channel SNR rather than timestep. 
In this scenario, $s$-dB channel is at earlier timestep than $t$-dB channel, but it is found $t < s$.
Given $T$ different channels in CSNRs $\{s_1, \dots, s_T\}$, this observation requires $s_1 > \dots > s_T$.
In summary, the decreasing CSNR order is required in forward diffusion process.

VDM requires $\text{VSNR}(i)=\alpha_i^2/\sigma_i^2$ to decrease over timesteps. Since $\text{VSNR}(i)=1/w_{s_i}^2$ and $w_{s_i}\propto 1/\sqrt{s_i}$, this is equivalent to the decreasing CSNR order established above.

\subsection{Belief Propagation in Reverse Process}
Unlike the noise-prediction framework \cite{bennatan2018deep} that requires syndrome as extra input, BP directly predicts $\mathbf{x}$ from LLR alone. We augment BP with a lightweight neural network to boost performance while retaining its low complexity.

Existing neural belief propagation (NBP) decoding algorithms \cite{nachmani2016learning} learn weight parameters on all message inputs in Eq. \ref{eq:bp}.
As found in \cite{liao2021doubly}, such design can be over-parameterization given the similarity between $s_v$ and $u_{v\rightarrow c}$. 
It can be noticed that the difference between $s_v$ and $u_{v\rightarrow c}$ is simply $u_{c\rightarrow v}$.
Therefore, learning neural parameters on message $u_{c\rightarrow v}$ can be sufficient for decoding:
\begin{equation}
\label{eq:res}
\begin{aligned}
s_v &= u_{v\rightarrow c} + u_{c\rightarrow v}(\theta, u_{v\rightarrow c}) \\
&\implies \mathbf{x} \gets \mathbf{x} + f(\mathbf{w}, \mathbf{x}),
\end{aligned}
\end{equation}
where $\mathbf{w}$ is a shared layer weight, $f(\cdot)$ preserves the $u_{c\rightarrow v}$ message structure, and the first layer input is $l_v$ for all $v$.

One neural block has $N-K$ layers following Eq.~\ref{eq:res}; complexity can scale by stacking blocks (intra-scale) or adding reverse timesteps (inter-scale). Since additional timesteps incur no storage overhead, we favor the timestep direction for practical deployment.

For the reverse diffusion process, we use $\hat{\mathbf{x}}_\theta(\mathbf{z}_t)$ to represent the model prediction given input $\mathbf{z}_t$ at timestep $t$. 
The related transition probability can be expressed as: 
\begin{equation}
\begin{aligned}
p(\mathbf{z}_s|\mathbf{z}_t) &= q(\mathbf{z}_s|\mathbf{z}_t,\mathbf{x}=\hat{\mathbf{x}}_\theta(\mathbf{z}_t)) \\
&= \mathcal{N}(\mathbf{\mu}_Q, \mathbf{\sigma}_Q) \\
&= \mathcal{N}\!\left(
    \tfrac{\alpha_{t|s}\sigma^2_s}{\sigma^2_t}\mathbf{z}_t
    + \tfrac{\alpha_s\sigma^2_{t|s}}{\sigma^2_t}\hat{\mathbf{x}}_\theta(\mathbf{z}_t;t),\;
    \tfrac{\sigma^2_{t|s}\sigma^2_s}{\sigma^2_t}\mathbf{I}\right) \\
&= \mathcal{N}\!\bigg(\mathbf{z}_t
    + \left(\tfrac{2}{w^2_s} - \tfrac{2}{w^2_t}\right)\hat{\mathbf{x}}_\theta(\mathbf{z}_t;t), \\
&\qquad\quad \left[\left(\tfrac{2}{w_s}\right)^{\!2}-\left(\tfrac{2}{w_t}\right)^{\!2}\right]\mathbf{I}\bigg).
\end{aligned}
\end{equation}
In addition, the purpose of channel coding is more focused on denoising than generation.
\cite{DBLP:conf/iclr/ChoukrounW23} choose to skip the noise addition step in the reverse process which is often found in DDPMs.
Besides, they also use number of parity checks in $\mathbf{H}$ as the maximum  reverse timestep.
We take a similar approach for reverse process except that we find their reverse timestep bound is loose because $N-K$ can often be unnecessarily large causing high complexity.
In practice, we limit our reverse process up to 20 timesteps which already achieves competitive results. 
The number of parity check errors is also applied to perform early stopping during the reverse process. 

\section{Experiments}

\begin{table*}[t]
\caption{$-\ln(\text{BER})$ of VCDC at different reverse timesteps (\textbf{higher is better}). CSNR: 4, 5, 6\,dB. ``Ours-$T$'' denotes our model at reverse timestep $T$.}
\label{tab:sampling}
\centering
\scriptsize
\begin{tabular}{ll|ccc|ccc|ccc|ccc}
\toprule
& & \multicolumn{3}{|c|}{Ours-1} & \multicolumn{3}{|c|}{Ours-5} & \multicolumn{3}{|c|}{Ours-10} & \multicolumn{3}{|c}{Ours-20} \\
Code & (N, K) & 4 & 5 & 6 & 4 & 5 & 6 & 4 & 5 & 6 & 4 & 5 & 6 \\
\midrule
LDPC& (121, 60) & 3.98 & 5.82 & 8.75 & 4.71 & 7.48 & 11.89 & 5.1 & 8.14 & 12.98 & 5.24 & 8.55 & 13.21 \\
LDPC &(121, 70) & 4.72 & 6.95 & 9.94 & 6.01 & 9.48 & 14.41 & 6.33 & 10.04 & 15.42 & 6.59 & 10.36 & 15.42 \\
LDPC &(121, 80) & 5.24 & 7.64 & 10.65 & 6.8 & 10.49 & 15.46 & 7.27 & 11.15 & 16.79 & 7.48 & 11.65 & 17.2 \\
LDPC &(49, 24) & 4.54 & 6.06 & 8.24 & 5.48 & 7.43 & 10.31 & 5.86 & 7.83 & 11.07 & 5.8 & 7.96 & 11.2 \\
Polar &(128, 64) & 2.88 & 3.28 & 3.74 & 3.74 & 4.84 & 6.12 & 4.84 & 6.62 & 9.09 & 6.46 & 9.31 & 13.11 \\
Polar &(128, 86) & 3.41 & 3.92 & 4.55 & 4.48 & 5.75 & 7.4 & 5.45 & 7.1 & 9.25 & 6.56 & 8.87 & 11.85 \\
Polar &(128, 96) & 3.66 & 4.22 & 4.93 & 4.5 & 5.84 & 7.61 & 5.46 & 7.7 & 10.4 & 6.35 & 8.86 & 12.33 \\
Polar &(64, 32) & 2.86 & 3.28 & 3.76 & 4.4 & 5.44 & 6.68 & 5.36 & 6.58 & 8.54 & 6.31 & 8.61 & 10.96 \\
Polar &(64, 48) & 3.76 & 4.48 & 5.32 & 4.8 & 6.25 & 8.1 & 5.63 & 7.56 & 9.84 & 6.14 & 7.95 & 10.61 \\
CCSDS& (128, 64) & 4.43 & 6.11 & 8.37 & 6.23 & 9.82 & 14.39 & 6.98 & 10.95 & 16.54 & 7.61 & 11.68 & 17.01 \\
MACKAY& (96, 48) & 4.67 & 6.07 & 7.96 & 6.39 & 9.12 & 12.06 & 7.26 & 10.38 & 13.76 & 7.59 & 11.04 & 14.45 \\
\bottomrule
\end{tabular}
\end{table*}

\begin{table*}[t]
\caption{$-\ln(\text{BER})$ comparison between models (\textbf{higher is better}). CSNR: 4, 5, 6\,dB.}
\label{tab:ber}
\centering
\scriptsize
\begin{tabular}{ll|lll|lll|lll|lll}
\toprule
& & \multicolumn{3}{|c|}{BP} & \multicolumn{3}{|c|}{HGN} & \multicolumn{3}{|c|}{DDECC} & \multicolumn{3}{|c}{Ours-20} \\
Code &(N, K) & 4 & 5 & 6 & 4 & 5 & 6 & 4 & 5 & 6 & 4 & 5 & 6 \\
\midrule
LDPC& (121, 60) & 4.82 & 7.21 & 10.87 & 5.22 & 8.29 & 13.0 & 4.48 & 6.95 & 10.65 & 5.24 & 8.55 & 13.21 \\
LDPC& (121, 70) & 5.88 & 8.76 & 13.04 & 6.39 & 9.81 & 14.04 & 5.41 & 8.22 & 12.22 & 6.59 & 10.36 & 15.42 \\
LDPC& (121, 80) & 6.66 & 9.82 & 13.98 & 6.95 & 10.68 & 15.8 & 6.12 & 9.38 & 13.25 & 7.48 & 11.65 & 17.2 \\
LDPC& (49, 24) & 5.3 & 7.28 & 9.88 & 5.76 & 7.9 & 11.17 & 5.27 & 7.38 & 10.23 & 5.8 & 7.96 & 11.2 \\
Polar& (128, 64) & 3.38 & 3.8 & 4.15 & 3.89 & 5.18 & 6.94 & 5.37 & 7.75 & 10.51 & 6.46 & 9.31 & 13.11 \\
Polar& (128, 86) & 3.8 & 4.19 & 4.62 & 4.57 & 6.18 & 8.27 & 5.61 & 7.76 & 10.42 & 6.56 & 8.87 & 11.85 \\
Polar& (128, 96) & 3.99 & 4.41 & 4.78 & 4.73 & 6.39 & 8.57 & 5.6 & 7.83 & 10.56 & 6.35 & 8.86 & 12.33 \\
Polar& (64, 32) & 3.52 & 4.04 & 4.48 & 4.25 & 5.49 & 7.02 & 5.99 & 8.16 & 10.9 & 6.31 & 8.61 & 10.96 \\
Polar& (64, 48) & 4.15 & 4.68 & 5.31 & 4.91 & 6.48 & 8.41 & 5.55 & 7.67 & 10.08 & 6.14 & 7.95 & 10.61 \\
CCSDS &(128, 64) & 6.55 & 9.65 & 13.78 & 6.99 & 10.57 & 15.27 & 5.79 & 8.48 & 12.24 & 7.61 & 11.68 & 17.01 \\
MACKAY &(96, 48) & 6.84 & 9.4 & 12.57 & 7.19 & 10.02 & 13.16 & 6.18 & 8.63 & 11.53 & 7.59 & 11.04 & 14.45 \\
\bottomrule
\end{tabular}
\end{table*}

We set CSNR from 4-dB to 6-dB as previous work \cite{DBLP:conf/iclr/ChoukrounW23} adopts and evaluate our method on different linear block codes, i.e., Polar Codes \cite{arikan2009channel}, Low-Density Parity Check (LDPC) codes \cite{gallager1962low}, Mackay codes and CCSDS codes, which are available on \cite{channelcodes}. 
Modern channel coding has stringent latency and model size requirement.
Although deep learning-based decoders often benefit from the power of increasing model size, the resulting deployment cost is not feasible in the case of channel coding, e.g., millisecond-level latency tolerance \cite{rico2020survey}.
We compare different models in their lightest setting over the bit error rate (BER).

To achieve maximum efficiency, we build our VDCD models using a single block neural network and set $T=20$ for reverse process. 
For model training, Adam optimizer \cite{DBLP:journals/corr/KingmaB14} is applied using learning rate 0.001 with 256 samples per batch and 20000 training iterations. 
Different from many diffusion models, the high efficiency of our model enables training and experiments on a CPU only platform, i.e., AMD EPYC 7402P 24-Core Processor. 
The training time varies between 1 to 3 hours for different codes. 

Comparison is made with the hyper graph neural network-based model \cite{nachmani2019hyper} that also maintains the belief propagation structure, which is referred as HGN. 
Their fastest models are configured with hidden dimension 32 and 5 hidden layers. 
We also compare with the state-of-the-art DDECC \cite{DBLP:conf/iclr/ChoukrounW23}, where their fastest models are configured with 2 self attention layers in hidden dimension 32. 
The traditional belief propagation algorithm running 5 iterations is also listed as the baseline. 

\subsection{Bit Error Rate}
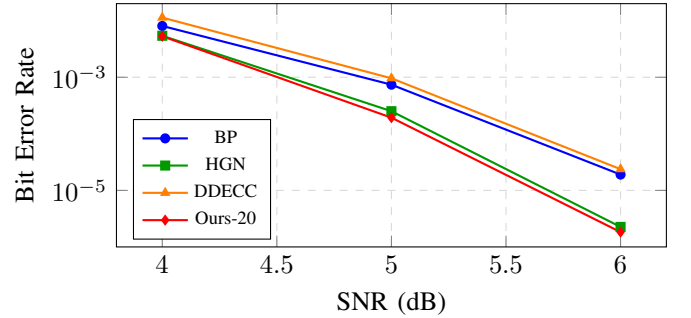
\begin{figure}[tb]
\centering
\begin{tikzpicture}
\begin{semilogyaxis}[
    width=\columnwidth,
    height=4.8cm,
    xlabel={SNR (dB)},
    ylabel={Bit Error Rate},
    grid=both,
    grid style={dashed, gray!30},
    legend style={at={(0.03,0.03)}, anchor=south west, font=\scriptsize},
    mark size=1.5pt,
    xmin=3.8, xmax=6.2,
    ymin=1e-6, ymax=2e-2, 
]
\addplot[mark=*, blue, thick] coordinates {(4, 0.008066) (5, 0.000739) (6, 0.00001902)}; \addlegendentry{BP}
\addplot[mark=square*, green!60!black, thick] coordinates {(4, 0.005407) (5, 0.000251) (6, 0.00000226)}; \addlegendentry{HGN}
\addplot[mark=triangle*, orange, thick] coordinates {(4, 0.011333) (5, 0.000958) (6, 0.0000237)}; \addlegendentry{DDECC}
\addplot[mark=diamond*, red, thick] coordinates {(4, 0.005299) (5, 0.000193) (6, 0.000001832)}; \addlegendentry{Ours-20}
\end{semilogyaxis}
\end{tikzpicture}
\caption{Standard BER curves for LDPC (121, 60).}
\label{fig:ber}
\end{figure}

We report negative log BER (higher is better) so that large differences in magnitude are directly comparable; evaluation halts once at least 100 error samples are detected per code. We also provide standard BER curves (e.g., Fig.~\ref{fig:ber}) for better readability.

Table~\ref{tab:sampling} shows BER improving monotonically with more reverse timesteps, with gains slowing toward convergence at step 20. This confirms the effectiveness of the reverse process; early stopping (halting when parity-check errors reach zero) contributes to consistent BER improvement across timesteps.

Table~\ref{tab:ber} compares models at their lightest configurations. BP is a surprisingly strong baseline—both our Ours-1 and DDECC can fall below it on some codes, while HGN consistently surpasses it. The observation that DDECC underperforms BP in certain cases (e.g., LDPC-121,60) is expected: DDECC's self-attention fails to implicitly learn the bipartite graph under strict parameter constraints. In contrast, VCDC directly integrates the exact parity-check matrix, ensuring robustness at minimal scales. Overall, our Ours-20 achieves the best BER across all codes and SNR levels. Deep learning models show larger gains at $5\!\rightarrow\!6$ dB than at $4\!\rightarrow\!5$ dB, a pattern that does not hold consistently for BP on Polar codes.

\begin{table}[tb]
\caption{Complexity Comparison (LDPC 121,60 lightest setting)}
\label{tab:complexity}
\centering
\scriptsize
\begin{tabular}{lrr}
\toprule
Model & Total FLOPs & Model Size \\
\midrule
BP & 316.4 K & 0 B \\
HGN & 1.6 G & 1.6 MB \\
DDECC-Max & 140.3 G & 226.3 KB \\
Ours-20 & \textbf{377.6 K} & \textbf{264.0 B} \\
\bottomrule
\end{tabular}
\end{table}

\subsection{Complexity and Latency}
To address computational and latency constraints, Table~\ref{tab:complexity} provides a quantitative comparison under the lightest setting for a representative code, LDPC (121,60). While wall-clock decoding latency depends heavily on hardware, FLOPs serve as a deterministic proxy. Our VCDC achieves up to 5 orders of magnitude reduction in FLOPs compared to the strong baseline DDECC. Moreover, VCDC's minimal storage (264 Bytes) translates to fewer than 100 learnable parameters. Finally, the early stopping mechanism reduces the actual average reverse steps well below the theoretical maximum ($T=20$), further minimizing practical average decoding latency.

\section{Conclusions}
This paper proposes a diffusion based decoder for channel coding.
Driven by the strict requirements of low latency and high reliability, our model is designed to leverage traditional belief propagation and modern diffusion frameworks. 
The method reformulates the AWGN channel as forward diffusion process in the VDM framework and builds a neural network architecture on top of belief propagation. 
Experiments show our design achieves the best decoding performance with significantly reduced computation and storage cost compared to the state-of-the-art diffusion based decoder.
These results exhibit the great potential of deploying diffusion based decoder in reality. 

\section*{Acknowledgement}
This work was financially supported by the National Key R\&D Program of China (Grant No. 2024YFA1211400).

\bibliographystyle{IEEEtran}
\bibliography{ref}

\end{document}